\documentclass[conference,]{IEEEtran}
\usepackage{cite}
\usepackage{amsmath,amssymb,amsfonts}
\usepackage{pifont}
\usepackage[%
    hyperfootnotes=false,%
    colorlinks=false,%
    pdfborder={0 0 0},%
    pdftitle={NISTT: A Non-Intrusive SystemC-TLM 2.0 Tracing Tool},%
    pdfauthor={Nils Bosbach et al.},%
    pdfkeywords={SystemC, TLM, ESL, LD\textunderscore PRELOAD},%
    ]{hyperref}
\usepackage{graphicx}
\usepackage{textcomp}
\usepackage{xcolor}
\usepackage{placeins}
\usepackage{xspace}
\usepackage[printonlyused,nolist]{acronym}
\usepackage[capitalise,nameinlink]{cleveref}
\usepackage[binary-units]{siunitx}
\usepackage{multirow}
\usepackage{booktabs}
\usepackage{stfloats}
\usepackage{fontawesome}
\usepackage{hyperref}
\usepackage{arydshln}
\ifCLASSOPTIONcompsoc
    \usepackage[caption=false,font=normalsize,labelfont=sf,textfont=sf]{subfig}
\else
    \usepackage[caption=false,font=footnotesize]{subfig}
\fi

\usepackage{tikz}
\usepackage{pgfplots}
\usetikzlibrary{
    positioning,
    shapes,
    arrows,
    pgfplots.external,
    pgfplots.groupplots,
    pgfplots.dateplot,
    fit}

\pgfplotsset{compat=newest}

\usepackage[pages=some]{background}

\def\BibTeX{{\rm B\kern-.05em{\sc i\kern-.025em b}\kern-.08em
    T\kern-.1667em\lower.7ex\hbox{E}\kern-.125emX}}

\newcommand*\circled[1]{\tikz[baseline=(char.base)]{
            \node[shape=circle,draw,fill=white,inner sep=1pt] (char) {#1};}}

\DeclareMathOperator{\rtf}{RTF}

\makeatletter
    \newcommand{\linebreakand}{%
      \end{@IEEEauthorhalign}
      \hfill\mbox{}\par
      \mbox{}\hfill\begin{@IEEEauthorhalign}
    }
\makeatother

\definecolor{rwth}   {RGB}{  0  84 159}
\definecolor{rwth-75}{RGB}{ 64 127 183}
\definecolor{rwth-50}{RGB}{142 186 229}
\definecolor{rwth-25}{RGB}{199 221 242}
\definecolor{rwth-10}{RGB}{232 241 250}

\definecolor{black}   {RGB}{  0   0   0}
\definecolor{black-75}{RGB}{100 101 103}
\definecolor{black-50}{RGB}{156 158 159}
\definecolor{black-25}{RGB}{207 209 210}
\definecolor{black-10}{RGB}{236 237 237}

\definecolor{magenta}   {RGB}{227   0 102}
\definecolor{magenta-75}{RGB}{233  96 136}
\definecolor{magenta-50}{RGB}{241 158 177}
\definecolor{magenta-25}{RGB}{249 210 218}
\definecolor{magenta-10}{RGB}{253 238 240}

\definecolor{yellow}   {RGB}{255 237   0}
\definecolor{yellow-75}{RGB}{255 240  85}
\definecolor{yellow-50}{RGB}{255 245 155}
\definecolor{yellow-25}{RGB}{255 250 209}
\definecolor{yellow-10}{RGB}{255 253 238}

\definecolor{petrol}   {RGB}{  0  97 101}
\definecolor{petrol-75}{RGB}{ 45 127 131}
\definecolor{petrol-50}{RGB}{125 164 167}
\definecolor{petrol-25}{RGB}{191 208 209}
\definecolor{petrol-10}{RGB}{230 236 236}

\definecolor{turkis}   {RGB}{  0 152 161}
\definecolor{turkis-75}{RGB}{  0 177 183}
\definecolor{turkis-50}{RGB}{137 204 207}
\definecolor{turkis-25}{RGB}{202 231 231}
\definecolor{turkis-10}{RGB}{235 246 246}

\definecolor{grun}   {RGB}{ 87 171  39}
\definecolor{grun-75}{RGB}{141 192  96}
\definecolor{grun-50}{RGB}{184 214 152}
\definecolor{grun-25}{RGB}{221 235 206}
\definecolor{grun-10}{RGB}{242 247 236}

\definecolor{maigrun}   {RGB}{189 205   0}
\definecolor{maigrun-75}{RGB}{208 217  92}
\definecolor{maigrun-50}{RGB}{224 230 154}
\definecolor{maigrun-25}{RGB}{240 243 208}
\definecolor{maigrun-10}{RGB}{249 250 237}

\definecolor{orange}   {RGB}{246 168   0}
\definecolor{orange-75}{RGB}{250 190  80}
\definecolor{orange-50}{RGB}{253 212 143}
\definecolor{orange-25}{RGB}{254 234 201}
\definecolor{orange-10}{RGB}{255 247 234}

\definecolor{rot}   {RGB}{204   7  30}
\definecolor{rot-75}{RGB}{216  92  65}
\definecolor{rot-50}{RGB}{230 150 121}
\definecolor{rot-25}{RGB}{243 205 187}
\definecolor{rot-10}{RGB}{250 235 227}

\definecolor{bordeaux}   {RGB}{161  16  53}
\definecolor{bordeaux-75}{RGB}{182  82  86}
\definecolor{bordeaux-50}{RGB}{205 139 135}
\definecolor{bordeaux-25}{RGB}{229 197 192}
\definecolor{bordeaux-10}{RGB}{245 232 229}

\definecolor{lila}   {RGB}{122 111 172}
\definecolor{lila-75}{RGB}{155 145 193}
\definecolor{lila-50}{RGB}{188 181 215}
\definecolor{lila-25}{RGB}{222 218 235}
\definecolor{lila-10}{RGB}{242 240 247}

\definecolor{violett}   {RGB}{ 97  33  88}
\definecolor{violett-75}{RGB}{131  78 117}
\definecolor{violett-50}{RGB}{168 133 158}
\definecolor{violett-25}{RGB}{210 192 205}
\definecolor{violett-10}{RGB}{237 229 234}

\definecolor{rwth}   {RGB}{  0  84 159}
\definecolor{rwth-75}{RGB}{ 64 127 183}
\definecolor{rwth-50}{RGB}{142 186 229}
\definecolor{rwth-25}{RGB}{199 221 242}
\definecolor{rwth-10}{RGB}{232 241 250}

\definecolor{black}   {RGB}{  0   0   0}
\definecolor{black-75}{RGB}{100 101 103}
\definecolor{black-50}{RGB}{156 158 159}
\definecolor{black-25}{RGB}{207 209 210}
\definecolor{black-10}{RGB}{236 237 237}

\definecolor{magenta}   {RGB}{227   0 102}
\definecolor{magenta-75}{RGB}{233  96 136}
\definecolor{magenta-50}{RGB}{241 158 177}
\definecolor{magenta-25}{RGB}{249 210 218}
\definecolor{magenta-10}{RGB}{253 238 240}

\definecolor{yellow}   {RGB}{255 237   0}
\definecolor{yellow-75}{RGB}{255 240  85}
\definecolor{yellow-50}{RGB}{255 245 155}
\definecolor{yellow-25}{RGB}{255 250 209}
\definecolor{yellow-10}{RGB}{255 253 238}

\definecolor{petrol}   {RGB}{  0  97 101}
\definecolor{petrol-75}{RGB}{ 45 127 131}
\definecolor{petrol-50}{RGB}{125 164 167}
\definecolor{petrol-25}{RGB}{191 208 209}
\definecolor{petrol-10}{RGB}{230 236 236}

\definecolor{turkis}   {RGB}{  0 152 161}
\definecolor{turkis-75}{RGB}{  0 177 183}
\definecolor{turkis-50}{RGB}{137 204 207}
\definecolor{turkis-25}{RGB}{202 231 231}
\definecolor{turkis-10}{RGB}{235 246 246}

\definecolor{grun}   {RGB}{ 87 171  39}
\definecolor{grun-75}{RGB}{141 192  96}
\definecolor{grun-50}{RGB}{184 214 152}
\definecolor{grun-25}{RGB}{221 235 206}
\definecolor{grun-10}{RGB}{242 247 236}

\definecolor{maigrun}   {RGB}{189 205   0}
\definecolor{maigrun-75}{RGB}{208 217  92}
\definecolor{maigrun-50}{RGB}{224 230 154}
\definecolor{maigrun-25}{RGB}{240 243 208}
\definecolor{maigrun-10}{RGB}{249 250 237}

\definecolor{orange}   {RGB}{246 168   0}
\definecolor{orange-75}{RGB}{250 190  80}
\definecolor{orange-50}{RGB}{253 212 143}
\definecolor{orange-25}{RGB}{254 234 201}
\definecolor{orange-10}{RGB}{255 247 234}

\definecolor{rot}   {RGB}{204   7  30}
\definecolor{rot-75}{RGB}{216  92  65}
\definecolor{rot-50}{RGB}{230 150 121}
\definecolor{rot-25}{RGB}{243 205 187}
\definecolor{rot-10}{RGB}{250 235 227}

\definecolor{bordeaux}   {RGB}{161  16  53}
\definecolor{bordeaux-75}{RGB}{182  82  86}
\definecolor{bordeaux-50}{RGB}{205 139 135}
\definecolor{bordeaux-25}{RGB}{229 197 192}
\definecolor{bordeaux-10}{RGB}{245 232 229}

\definecolor{lila}   {RGB}{122 111 172}
\definecolor{lila-75}{RGB}{155 145 193}
\definecolor{lila-50}{RGB}{188 181 215}
\definecolor{lila-25}{RGB}{222 218 235}
\definecolor{lila-10}{RGB}{242 240 247}

\definecolor{violett}   {RGB}{ 97  33  88}
\definecolor{violett-75}{RGB}{131  78 117}
\definecolor{violett-50}{RGB}{168 133 158}
\definecolor{violett-25}{RGB}{210 192 205}
\definecolor{violett-10}{RGB}{237 229 234}

\newcommand{\ldpreload}{\texttt{LD\_PRELOAD}\xspace}
\newcommand{\nistt}{NISTT\xspace}

\begin{document}
\title{NISTT: A Non-Intrusive SystemC-TLM~2.0\\Tracing Tool}

\author{
    \IEEEauthorblockN{Nils Bosbach,
        Jan Moritz Joseph, and
        Rainer Leupers}
    \IEEEauthorblockA{\textit{Institute for Communication Technologies and Embedded Systems}\\
        \textit{RWTH Aachen University}\\
        Aachen, Germany\\
        \{bosbach, joseph, leupers\}@ice.rwth-aachen.de}
    \and
    \IEEEauthorblockN{Lukas Jünger}
    \IEEEauthorblockA{\textit{MachineWare GmbH}\\
        Aachen, Germany\\
        lukas@mwa.re}
    }

\IEEEoverridecommandlockouts
\IEEEpubid{\makebox[\columnwidth]{978-1-6654-9005-4/22/\$31.00~\copyright2022 IEEE\hfill}
\hspace{\columnsep}\makebox[\columnwidth]{ }}
\newcommand\copyrighttext{%
  \footnotesize \textcopyright 2022 IEEE. Personal use of this material is permitted. Permission from IEEE must be obtained for all other uses, including reprinting/republishing this material for advertising or promotional purposes, collecting new collected works for resale or redistribution to servers or lists, or reuse of any copyrighted component of this work in other works.}

\newcommand\copyrightnotice{%
    \backgroundsetup{opacity=1, scale=1, angle=0, contents={
            \color{black}%
            \begin{tikzpicture}[remember picture,overlay]%
                \node[anchor=south,yshift=10pt] at (current page.south) {\fbox{\parbox{\dimexpr0.75\textwidth-\fboxsep-\fboxrule\relax}{\copyrighttext}}};
                \node[anchor=north,yshift=-10pt,text=gray] at (current page.north) {\shortstack[c]{PREPRINT - accepted by 30th IFIP/IEEE International Conference on Very Large Scale Integration 2022 (VLSI-SoC 2022)\\DOI: \href{https://doi.org/10.1109/VLSI-SoC54400.2022.9939578}{10.1109/VLSI-SoC54400.2022.9939578}}};
            \end{tikzpicture}%
        }%
    }%
    \BgThispage%
}

\maketitle
\copyrightnotice

\IEEEpubidadjcol

\begin{acronym}[AAPCS64]
    \acro{aapcs64}[AAPCS64]{Procedure Call Standard for the \acl{aarch64}}
    \acro{aarch64}[AArch64]{\acs{arm} 64-bit Architecture}
    \acro{ai}[AI]{Artificial Intelligence}
    \acro{aiba}[AIBA]{An Automated Intra-cycle Behavioral Analysis for SystemC-based design exploration}
    \acro{amd}[AMD]{Advanced Micro Devices}
    \acro{api}[API]{Application Programming Interface}
    \acro{arm}[ARM]{Advanced \acs{risc} Machines}
    \acro{ast}[AST]{Abstract Syntax Tree}
    \acro{avp64}[AVP64]{An \acs{arm}v8 \acl{vp}}
    \acro{bl}[BL]{Branch-With-Link}
    \acro{bp}[BP]{breakpoint}
    \acro{cpu}[CPU]{Central Processing Unit}
    \acro{crc}[CRC]{Cyclic Redundancy Check}
    \acro{csv}[CSV]{Character-Separated Values}
    \acro{db}[DB]{database}
    \acro{dbms}[DBMS]{\Acl{db} Management System}
    \acro{ddr}[DDR]{Double Data Rate}
    \acro{des}[DES]{Discrete Event Simulation}
    \acro{dla}[DLA]{Deep Learning Accelerator}
    \acro{dmi}[DMI]{Direct Memory Interface}
    \acro{ds}[DS]{Developer Studio}
    \acro{dwarf}[DWARF]{Debugging With Arbitrary Record Formats}
    \acro{eda}[EDA]{Electronic Design Automation}
    \acro{elf}[ELF]{Executable and Linkable Format}
    \acro{esa}[ESA]{European Space Agency}
    \acro{esl}[ESL]{Electronic System Level}
    \acro{fd}[fd]{file descriptor}
    \acro{fig}[FIG]{Fault Injection in glibc}
    \acro{fp}[FP]{Frame Pointer}
    \acro{fpga}[FPGA]{Field Programmable Gate Array}
    \acro{fvp}[FVP]{Fixed Virtual Platform}
    \acro{gcc}[GCC]{GNU Compiler Collection}
    \acro{gdb}[GDB]{GNU Debugger}
    \acro{gic}[GIC]{Generic Interrupt Controller}
    \acro{got}[GOT]{Global Offset Table}
    \acro{gui}[GUI]{Graphical User Interface}
    \acro{hw}[HW]{Hardware}
    \acro{ibm}[IBM]{International Business Machines}
    \acro{id}[ID]{identifier}
    \acro{ieee}[IEEE]{Institute of Electrical and Electronics Engineers}
    \acro{io}[I/O]{Input/Output}
    \acro{ir}[IR]{Intermediate Representation}
    \acro{irq}[IRQ]{Interrupt Request}
    \acro{iss}[ISS]{Instruction-Set Simulator}
    \acro{lr}[LR]{Link Register}
    \acro{miso}[MISO]{Master Input Slave Output}
    \acro{mnist}[MNIST]{Modified National Institute of Standards and Technology}
    \acro{mosi}[MOSI]{Master Output Slave Input}
    \acro{nvdla}[NVDLA]{NVIDIA \acl*{dla}}
    \acro{os}[OS]{Operating System}
    \acrodefplural{os}[OS's]{Operating Systems}
    \acro{pc}[PC]{Program Counter}
    \acro{pid}[PID]{process ID}
    \acro{pccts}[PCCTS]{Purdue Compiler Construction Tool Set}
    \acro{pci}[PCI]{Peripheral Component Interconnect}
    \acro{plt}[PLT]{Procedure Linkage Table}
    \acro{pmu}[PMU]{Performance Monitoring Unit}
    \acro{qemu}[QEMU]{Quick Emulator}
    \acro{ram}[RAM]{Random-Access Memory}
    \acro{risc}[RISC]{Reduced Instruction Set Computer}
    \acro{rtf}[$\rtf$]{Real-Time Factor}
    \acro{rtl}[RTL]{Register-Transfer Level}
    \acro{rtos}[RTOS]{real-time operating system}
    \acro{sata}[SATA]{Serial AT Attachment}
    \acro{sd}[SD]{Secure Digital}
    \acro{sdhci}[SDHCI]{\acs{sd} Host Controller Interface}
    \acro{soc}[SoC]{System-on-a-chip}
    \acrodefplural{soc}[SoCs]{Systems-on-a-chip}
    \acro{sp}[SP]{Stack Pointer}
    \acro{spi}[SPI]{Serial Peripheral Interface}
    \acro{sql}[SQL]{Structured Query Language}
    \acro{ssd}[SSD]{Solid State Drive}
    \acro{sw}[SW]{Software}
    \acro{tcp}[TCP]{Transmission Control Protocol}
    \acro{tlm}[TLM]{Transaction-level Modeling}
    \acro{uart}[UART]{Universal Asynchronous Receiver / Transmitter}
    \acro{unix}[UNIX]{Uniplexed Information and Computing Service}
    \acro{vcd}[VCD]{Value Change Dump}
    \acro{vcml}[VCML]{Virtual Components Modeling Library}
    \acro{viper}[VIPER]{Virtual Platform Explorer}
    \acro{vp}[VP]{Virtual Platform}
    \acro{wfi}[WFI]{Wait For Interrupt}
\end{acronym}

\begin{abstract}
    The increasing complexity of systems-on-a-chip requires the continuous development of electronic design automation tools.
Nowadays, the simulation of systems-on-a-chip using virtual platforms is common.
Virtual platforms enable hardware/software co-design to shorten the time to market, offer insights into the models, and allow debugging of the simulated hardware.
Profiling tools are required to improve the usability of virtual platforms.
During simulation, these tools capture data that are evaluated afterward.
Those data can reveal information about the simulation itself and the software executed on the platform.

This work presents the tracing tool \textit{\nistt} that can profile SystemC-TLM-2.0-based virtual platforms.
\nistt is implemented in a completely non-intrusive way.
That means no changes in the simulation are needed, the source code of the simulation is not required, and the traced simulation does not need to contain debug symbols.
The standardized SystemC application programming interface guarantees the compatibility of \nistt with other simulations.
The strengths of \nistt are demonstrated in a case study.
Here, \nistt is connected to a virtual platform and traces the boot process of Linux.
After the simulation, the database created by \nistt is evaluated, and the results are visualized.
Furthermore, the overhead of \nistt is quantified.
It is shown that \nistt has only a minor influence on the overall simulation performance.

\end{abstract}


\section{Introduction}
\label{sec:intro}
Nowadays, simulation is an important part of the design process of a \ac{soc}.
\acp{vp} can be developed in an early stage of the design process to serve as the enabler of \ac{hw}/\ac{sw} co-design.
A \ac{vp} is a simulation of a complete \ac{soc} that can execute target \ac{sw} without modification.
To support the development, standardized frameworks like SystemC-\acs{tlm}~2.0 are used \cite{ieee_standards_association_and_others_ieee_2012}.
SystemC defines standardized interfaces for models and their connection at the \ac{esl}.
The \ac{tlm} extension abstracts communications between models like memory accesses or interrupts to speed up the simulation.

Compared to real \ac{hw}, \acp{vp} have the benefit of allowing easy debugging and analysis.
Due to Moore's law and the related increasing complexity of \acp{soc}, simulations have also become more nested and complex.
This problem leads to performance issues in the simulation and the target \ac{sw}.
Additional tools are required to reveal which parts of the simulation or the target \ac{sw} need to be optimized to overcome those issues.
Those tools are called \textit{SystemC frontends}.
A distinction is made between \textit{static}, \textit{dynamic}, and \textit{hybrid} approaches.
Static approaches analyze the source code of the simulation and extract information without executing the simulation.
Dynamic techniques take the dynamic behavior of the simulation into account.
Modules that are created during runtime and the workload that is executed on the \ac{vp} are considered.
For that, the simulation is executed and analyzed.
Hybrid approaches combine the two techniques by analyzing both the static and the dynamic behavior.
Often, static analysis is used to search and annotate functions of interest.
Based on those annotations, the dynamic behavior is analyzed.
The static evaluation requires a tool that reads the source code of the simulation.
This tool can either be a parser or a compiler.

\begin{figure}[!t]
    \centering
    \tikzstyle{nodetype}  = [rectangle, rounded corners, minimum width=2cm, minimum height=.6cm, text centered, text width= 2cm, fill=rwth-25, font={\footnotesize}, draw=black]
\tikzstyle{nodetype2} = [nodetype, fill=bordeaux-25]
\tikzstyle{nodetype3} = [nodetype, fill=grun-25]
\tikzstyle{nodetype3} = [nodetype, fill=orange-25]
\tikzstyle{arrow}     = [thick,->,>=stealth]

\pgfdeclarelayer{background}
\pgfsetlayers{background,main} 

\begin{tikzpicture}
    \node[nodetype3]                                                        (vp)   {\acs{vp}};
    \node[nodetype2, right=of vp]                                           (snif) {\nistt};
    \node[nodetype3, right=of snif]                                         (sysc) {SystemC};
    \node[rectangle, below=of snif, yshift=+0.6cm, inner sep=0pt]           (db)   {\huge\faDatabase};
    \node[nodetype]                                           at (db-|sysc) (pp)   {Evaluation};
    \node[draw=none, above=of snif, yshift=-0.9cm]                          (sim)  {\footnotesize Simulation};

    \draw[arrow] (vp)   -- (snif);
    \draw[arrow] (snif) -- (db);
    \draw[arrow] (snif) -- (sysc);
    \draw[arrow] (db)   -- (pp);

    \begin{pgfonlayer}{background}
        \node[nodetype, fit={(vp)(snif)(sysc)(sim)}, minimum width=\linewidth] (sim_box) {};
    \end{pgfonlayer}

\end{tikzpicture}
    \vspace{-1em}
    \caption{\nistt approach.}
    \label{fig:nist-scheme-simple}
    \vspace{-1em}
\end{figure}

In this paper, we present \textit{\nistt} as a novel approach to tracing the behavior of SystemC-\acs{tlm}-2.0-based simulations.
The design goals of \nistt are:
\begin{itemize}
    \item Revealing insights of the simulation and the target \ac{sw}
    \item Tracing of pre-compiled \acp{vp} without debugging symbols
    \item No access to the source code of the simulation
    \item Capturing the dynamic behavior without a static analysis
\end{itemize}

As shown in \cref{fig:nist-scheme-simple}, \nistt is placed between the \ac{vp} and the SystemC library.
It can intercept function calls to the SystemC library to analyze data.
\nistt does not place any requirements on the simulation or the used toolchain.
The official SystemC library and the preferred toolchain can be used without changes.
The standardization of the SystemC \ac{api} guarantees compatibility.
Furthermore, \nistt is invisible to the simulation.
The simulation behavior is untouched.


\section{Related Work}
\label{sec:related-work}

\begin{table*}[!b]
    \vspace{-1em}
    \centering
    \caption{Overview of existing SystemC frontends.}
    \label{tab:sysc-frontend}
    \begin{tabular}{c c c c c c p{6cm}}
        \toprule
        \multirow{2}{*}{SystemC Frontend}        & \multicolumn{2}{c}{Analysis} & \multirow{2}{*}{Non-Intrusive} & \multicolumn{2}{c}{Works Without Access To} & \multirow{2}{*}{Used Tools}          \\ 
                                                 & Static       & Dynamic       &                                & Source Code   & Debug Symbols    &                                                 \\ \midrule
        SystemCXML \cite{berner_systemcxml_2005} & \checkmark   &               & \checkmark                     & \ding{55}     & \checkmark       & Doxygen                                         \\ \hdashline
        ParSysC \cite{fey_parsyc_2004}           & \checkmark   &               & \checkmark                     & \ding{55}     & \checkmark       & Parser                                          \\ \hdashline
        Genz et al. \cite{genz_overcoming_2009}  & \checkmark   & \checkmark    & \checkmark                     & \ding{55}     & \checkmark       & \acs{pccts}-based parser                        \\ \hdashline
        Pinapa \cite{moy_pinapa_2005}            & \checkmark   & \checkmark    & \ding{55}                      & \ding{55}     & \checkmark       & Modified \acs{gcc}, modified SystemC library    \\ \hdashline
        PinaVM \cite{marquet_pinavm_2010}        & \checkmark   & \checkmark    & \ding{55}                      & \ding{55}     & \checkmark       & LLVM, modified SystemC library                  \\ \hdashline
        SHaBE \cite{broeders_extracting_2011}    & \checkmark   & \checkmark    & \checkmark                     & \ding{55}     & \ding{55}        & \acs{gcc} plugin, \acs{gdb}                     \\ \hdashline
        AIBA \cite{goli_aiba_2016}               & \checkmark   & \checkmark    & \checkmark                     & \ding{55}     & \ding{55}        & \acs{gdb}                                       \\ \hdashline
        Quiny \cite{schubert_quiny_2007}         &              & \checkmark    & \ding{55}                      & \ding{55}     & \checkmark       & Modified SystemC library                        \\ \hdashline
        Scoot \cite{blanc_scoot_2008}            & \checkmark   &               & \ding{55}                      & \ding{55}     & \checkmark       & Custom C++ frontend, simplified SystemC library \\ \hdashline
        ReSp \cite{beltrame_resp_2009}           &              & \checkmark    & \checkmark                     & \ding{55}     & \checkmark       & Python wrapper                                  \\ \hdashline
        Stoppe et al. \cite{stoppe_data_2013}    & \checkmark   & \checkmark    & \checkmark                     & \ding{55}     & \ding{55}        & Debug symbols, SystemC \acs{api}                \\ \midrule
        \nistt                                   &              & \checkmark    & \checkmark                     & \checkmark    & \checkmark       & \texttt{LD\_PRELOAD}                            \\ \bottomrule
    \end{tabular}
\end{table*}

Over the past years, many SystemC frontends have been developed.
An overview of different frontends has been created by Marquet et al. in \cite{marquet_theoretical_2010}.
\cref{tab:sysc-frontend} shows an extended compilation of existing approaches.
As mentioned before, the approaches can be classified as \textit{static}, \textit{dynamic}, or \textit{hybrid}.

The static approaches are often based on a parser that analyzes the C++ source code of the simulation and derives information from the parsed output.
Different approaches for the implementation of the parser exist.
\textit{SystemCXML} \cite{berner_systemcxml_2005} uses Doxygen's C++ front end to parse the module code.
\textit{ParSysC} \cite{fey_parsyc_2004} uses a \ac{pccts}-based parser to convert the SystemC representation to a \ac{rtl} \ac{ir} to analyze the simulated system.
Genz et al. also developed a \ac{pccts}-based parser for the static analysis and a code generator that injects additional code into the simulation to evaluate runtime information \cite{genz_overcoming_2009}.

A problem that often occurs when SystemC parsers are used is the limitation to a subset of the SystemC language.
Therefore, compilers can be used instead of parsers to extract static information.
\textit{Pinapa} \cite{moy_pinapa_2005} uses a modified \ac{gcc} to get the abstract syntax tree of the simulation from which static information is extracted.
Dynamic information is extracted by executing the elaboration phase of the simulation, in which SystemC builds up the module hierarchy.
\textit{Pinapa} has been further developed to \textit{PinaVM}~\cite{marquet_pinavm_2010}.
\textit{PinaVM} uses the LLVM \ac{ir} to insert additional code into the simulation that is used to capture runtime information.

\textit{SHaBE} \cite{broeders_extracting_2011} and \textit{AIBA} \cite{goli_aiba_2016} use \ac{gdb} to debug the simulation during execution.
\textit{SHaBE} uses a \ac{gdb} plugin to extract data during the elaboration phase to build up the module hierarchy.
\textit{AIBA} creates a \ac{gdb} command file from a static analysis which controls \ac{gdb} during the execution to set breakpoints and store data.
This approach has been further developed to support the tracing of \ac{tlm} transactions \cite{goli_automated_2020}.
\textit{Quiny} \cite{schubert_quiny_2007} is a dynamic approach that uses a modified SystemC library that implements C++ operator overloading to retrieve information during runtime.
\textit{Scoot} \cite{blanc_scoot_2008} is a model extractor based on a custom C++ frontend that analyzes the source code using simplified SystemC header files to extract the module hierarchy, sensitivity list of the processes, and the port bindings.
\textit{ReSp} \cite{beltrame_resp_2009} adds Python wrappers to SystemC models which allow interaction during the simulation.
The approach proposed by Stoppe et al. \cite{stoppe_data_2013} uses the debug information of the compiled simulation executable and the SystemC \ac{api} to capture simulation-related data.

Most presented static approaches try to extract the module hierarchy and the connections between the modules.
This information does not include the runtime behavior of the simulation.
Furthermore, all approaches need access to the source code of the simulation to either directly analyze it or compile the simulation using customized tools.
That can be a drastic limitation, especially for industrial \acp{vp} where the source code is not distributed to the customer.
Another aspect that stood out is the need for an extensive static analysis most hybrid approaches use to configure their dynamic analysis.
It would be beneficial to perform the dynamic analysis without a preceding static one to keep the complexity of the tool as low as possible.
For those reasons, the idea for the development of \nistt is to create a tracing tool that works without a static analysis, does not require access to the source code of the simulation, and is as simple as possible.

\section{Proposed \nistt Approach}
\label{sec:proposed-approach}
The design idea behind \nistt is to create a tool that is capable of tracing an already compiled SystemC-\acs{tlm}-2.0-based simulation without accessing its source code or debug symbols.
These requirements increase the usability of the tracing tool compared to existing approaches.
To trace an already compiled simulation, \nistt intercepts the calls of the simulation to the shared SystemC library to extract data.
The interception of function calls without access to the source code can be done due to the standardized SystemC \ac{api}.
The \ldpreload feature of the Linux dynamic linker/loader, \textit{ld}, is used to perform this interception.
\textit{ld} is responsible for loading shared libraries that are needed by a program.
During runtime, \textit{ld} dynamically links function calls to those libraries.
The shared libraries that are needed by an executable are listed in the dynamic section of the compiled \ac{elf} file.
The Linux environment variable \ldpreload can be used to define paths of additional shared libraries that are loaded by \textit{ld}, regardless of whether they are required by the executable or not.
When the executable calls a function that is defined in a shared library, \textit{ld} needs to resolve that call.
The matching of the function to be called and the available functions is based on the function name.
In the case of a C program, that name corresponds to the name given by the programmer.
For C++ programs, a mangled name is used that is created by the compiler based on the name, return type, and parameter types of the function.

Preloading a library that contains a function with the same name as a function of a required shared library overrides the implementation of that function.
When the executable calls the function, \textit{ld} resolves the call.
\textit{ld} searches for an implementation of that function in the loaded shared objects.
The first function that is found is the one implemented in the preloaded library because it is loaded before the required libraries.
Thus, preloading can be used to intercept calls to a shared library by implementing a function with the same name in a preloaded library.

\subsection{Working Principle of \nistt}
\label{sec:proposed-approach:working-principle}
\nistt uses \ldpreload to intercept calls to the SystemC library and extract tracing information.
That enables interaction with the simulation without changing or accessing the source code of the simulation or dependent libraries. 
The only requirement is that the simulation must be dynamically linked against a SystemC-TLM-2.0-compatible library.

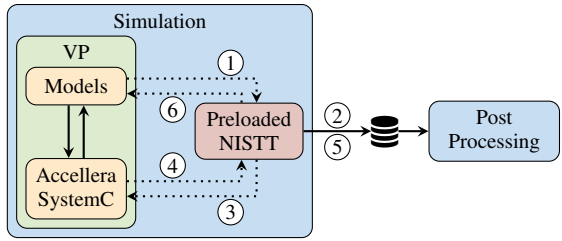
\begin{figure}[!tb]
    \centering
    \tikzstyle{nodetype} = [rectangle, rounded corners, minimum width=1cm, minimum height=.5cm, text centered, text width=1.1cm, draw=black, fill=white, font={\footnotesize}]
\tikzstyle{label}    = [draw=none, font={\footnotesize}, inner sep=0pt, outer sep=0pt]
\tikzstyle{icon}    = [draw=none, font={\large}, inner sep=0pt, outer sep=0pt]
\tikzstyle{arrow}    = [thick,->,>=stealth]

\pgfdeclarelayer{background}
\pgfdeclarelayer{middle}
\pgfsetlayers{background,middle,main} 

\begin{tikzpicture}[node distance=0.7cm and 0.1cm]
    \node[nodetype, fill=orange-25]                                                                  (mod)     {Models};
    \node[nodetype, fill=orange-25,   below=of mod]                                                  (sysc)    {Accellera SystemC};
    \node[label,                      above=of mod,                 yshift=-0.6cm]                   (vp_text) {VP};
    \node[nodetype, fill=bordeaux-25, right=of mod,  xshift=+0.8cm, yshift=-0.6cm, text width=1.2cm] (snif)    {Preloaded \nistt};
    \node[icon,                       right=of snif, xshift=+0.8cm]                                  (db)      {\faDatabase};
    \node[nodetype, fill=rwth-25,     right=of db,   xshift=+0.3cm,                text width=1.5cm] (pp)      {Post Processing};

    \draw[arrow]         ([xshift=-0.1cm]  mod.south) --                                                                  ([xshift=-0.1cm ] sysc.north);
    \draw[arrow]         ([xshift=+0.1cm] sysc.north) --                                                                  ([xshift=+0.1cm]   mod.south);
    \draw[arrow, dotted] ([yshift=+0.1cm]  mod.east)  -|  node[label, solid, above, pos=0.4, inner sep=1pt] {\circled{1}} ([xshift=+0.1cm]  snif.north);
    \draw[arrow, dotted] ([xshift=+0.1cm] snif.south) |-  node[label, solid, below, pos=0.6, inner sep=1pt] {\circled{3}} ([yshift=-0.1cm]  sysc.east);
    \draw[arrow, dotted] ([yshift=+0.1cm] sysc.east)  -|  node[label, solid, above, pos=0.2, inner sep=1pt] {\circled{4}} ([xshift=-0.1cm]  snif.south);
    \draw[arrow, dotted] ([xshift=-0.1cm] snif.north) |-  node[label, solid, below, pos=0.8, inner sep=1pt] {\circled{6}} ([yshift=-0.1cm]   mod.east);
    \draw[arrow]         (                snif.east)  --  node[label, solid, above,          inner sep=1pt] {\circled{2}} node[label, solid, below, inner sep=1pt] {\circled{5}} (db.west);
    \draw[arrow]         (                  db.east)  --                                                                  (                   pp.west);

    \begin{pgfonlayer}{middle}
        \node[nodetype, fit={(mod)(sysc)(vp_text)}, fill=grun-25] (vp_box) {};
    \end{pgfonlayer}

    \node[label, fit={(vp_box)(snif)}]                (sim_grp) {};
    \node[label, above=of sim_grp, yshift=-0.6cm]     (sim_lbl) {Simulation};

    \begin{pgfonlayer}{background}
        \node[nodetype, fit={(sim_grp)(sim_lbl)}, fill=rwth-25] (sim_box) {};
    \end{pgfonlayer}
\end{tikzpicture}
    \caption{\nistt working principle.}
    \label{fig:nist-scheme}
    \vspace{-1em}
\end{figure}

The working principle of \nistt is shown in \cref{fig:nist-scheme}.
\nistt is a library that needs to be preloaded to a SystemC-\acs{tlm}-2.0-based simulation using \ldpreload.
The library implements SystemC functions whose calls and passed data should be traced.
When the simulation calls such a function, \textit{ld} links the call to the \nistt implementation \circled{1}.
\nistt can then access and evaluate the passed parameters and store a data point in a database \circled{2}.
The original SystemC function is called afterward to keep the simulation behavior unchanged \circled{3}.
For that, the \ac{api} of the Linux dynamic linker/loader is used. 
After the original SystemC function returns \circled{4}, a second data point can be stored in the database \circled{5}.
Then, the \nistt function returns \circled{6}.
When the simulation calls a SystemC function that is not implemented in \nistt, \textit{dl} directly forwards the call to the original library, as shown by the solid arrows on the left.
%

\subsection{Traced SystemC Functions}
\label{sec:proposed-approach:traced-sysc-fcn}
\nistt can trace the following simulation properties:
\begin{itemize}
    \item SystemC process/coroutine scheduling
    \item Quantum duration of processes
    \item Processes waiting on event notifications 
    \item Notification of events
    \item Course of simulation time and real-time
\end{itemize}

\nistt overrides multiple variants of the SystemC function \texttt{wait} to intercept its calls.
\texttt{wait} can be used inside SystemC threads to suspend the execution of the thread in a non-preemptive way and resume it at a later point in time.
Parameters can be passed to the \texttt{wait} function to specify when the SystemC scheduler should resume the execution of the thread.
One variant of the \texttt{wait} function accepts an amount of simulation time that needs to pass until the thread is resumed.
Another implementation gets a SystemC event or a collection of SystemC events as a parameter.
In that case, the thread is resumed once the events have been notified.
When an overridden \texttt{wait} function of \nistt is called, the name of the calling SystemC process and simulation time/real-time timestamps are stored in a database along with information on the duration of the suspension.
Depending on the parameters of the \texttt{wait} function, that information is either the amount of simulation time that should be waited or the name of the event that needs to be notified.

The data stored on \texttt{wait} calls can be used for various analyses.
For instance, they provide information on the scheduling of SystemC threads.
SystemC threads are coroutines that use the \texttt{wait} function to suspend their execution by calling the scheduler.
During an intercepted \texttt{wait} call, the first data point is stored in the database when the execution of a thread is suspended \circled{2}.
Before the thread is resumed, a second data point is stored \circled{5}.
To trace the first entry into a thread, \nistt also intercepts calls to the function \texttt{sc\_thread\_cor\_fn}.
This function is used to invoke a coroutine.

Another property that can be derived from \texttt{wait} calls is the quantum duration.
In loosely-timed SystemC-\acs{tlm}~2.0 simulations, the concept of temporal decoupling is used.
Thereby, the simulation performance is increased by reducing the temporal accuracy.
SystemC threads are allowed to run ahead of the global simulation time to decrease the number of synchronizations.
They keep the elapsed time since their last synchronization with the global simulation time as their \textit{local time}.
When the local time of a thread exceeds a limit or an operation that requires high accuracy should be executed, the thread needs to synchronize with the global simulation time.
This synchronization is done by calling the \texttt{wait} function with the difference in time as a parameter.
This difference in time is called \textit{quantum}.
The used quantum durations of the process provide information on the simulation performance.
In general, large quantum durations are targeted because they increase the decoupling and thereby accelerate the simulation.

Apart from the \texttt{wait} function that gets a time as a parameter, SystemC also offers a function to suspend the execution of a thread until a specified event is notified.
This function can, e.g., be used in the implementation of a CPU model to implement a \ac{wfi} instruction.
To emulate the execution of \ac{wfi} on a \ac{vp}, \texttt{wait} can be called with the interrupt event as parameter.
Besides tracing threads that are waiting on notifications of events, \nistt can also trace the notification itself.
For that, the SystemC function \texttt{notify} of the class \texttt{sc\_event} is implemented in \nistt to intercept calls.
Depending on the parameters, the event is directly notified, or the notification is delayed by the specified amount of simulation time.

Every data point that is stored in the database contains timestamps of the current simulation time and the elapsed real-time since the beginning of the simulation.
Those two timestamps can be used to put the simulation time $t_{sim}$ in relation to the time needed for the simulation, the real-time $t_{real}$.
The \ac{rtf} can be calculated to measure the simulation performance.
\begin{equation*}
    \rtf = \frac{\Delta t_{sim}}{\Delta t_{real}} \label{equ:real-time-factor}
\end{equation*}

The data stored in the database can be visualized and evaluated in a post-processing step.
Visualizations using Python and Matplotlib \cite{hunter_matplotlib_2007} are presented in the next chapter.

In the future, additional SystemC functions can be implemented in \nistt to extend the functionality.
However, there are some limitations.
Inlined SystemC functions and methods defined in template classes cannot be overridden using preloading.
That is because they are directly compiled into the executable that uses those functions and therefore not stored in the shared library.

\section{Experimental Evaluation}
\label{sec:exp-eval}
\nistt is used to profile the boot process of a \ac{vp} in a case study to demonstrate the functionality of the tool.
The tracing overhead is measured and compared to an intrusive implementation.
As the profiled target, the \ac{vcml}-based \cite{weinstock_virtual_2022} \ac{vp} \acs{avp64} \cite{junger_fast_2019} is used.
Since \nistt is implemented in a non-intrusive way, no changes in SystemC, \ac{vcml}, or \acs{avp64} are needed.

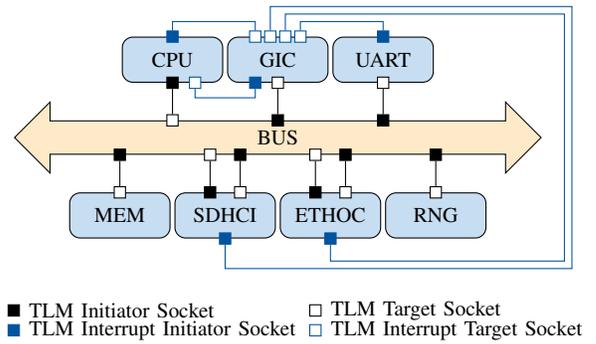
\begin{figure}[!t]
    \centering
    \tikzstyle{nodetype} = [rectangle, rounded corners, minimum width=1.1cm, minimum height=.6cm, text centered, text width=1.1cm, draw=black, fill=rwth-25, font={\footnotesize}]
\tikzstyle{isock}    = [rectangle, minimum width=.15cm, minimum height=.15cm, draw=black, fill=black, inner sep=0pt]
\tikzstyle{tsock}    = [isock, fill=white]
\tikzstyle{isocki}   = [isock, fill=rwth, draw=rwth]
\tikzstyle{tsocki}   = [isocki, fill=white]

\begin{tikzpicture}[node distance=.5cm and .5cm]
    \node[double arrow, draw, minimum height=7cm, fill=orange-25, font={\footnotesize}]    (bus)              {BUS};
    
    \node[nodetype, above=of bus.north, xshift=-1.4cm]                                     (cpu)              {CPU};
    \node[isock]                                       at (cpu.south)                      (cpu_sock)         {};
    \node[isocki]                                      at (cpu.north)                      (cpu_socki)        {};
    \node[tsocki]                                      at ([xshift=+0.3cm]cpu.south)       (cpu_sockit)       {};
    \node[tsock]                                       at (cpu_sock.center|-bus.north)     (bus_cpu_sock)     {};
    \draw (cpu_sock)      -- (bus_cpu_sock);

    \node[nodetype, above=of bus.north, xshift=+0.0cm]                                     (gic)              {GIC};
    \node[tsock]                                       at (gic.south)                      (gic_sock)         {};
    \node[isock]                                       at (gic_sock.center|-bus.north)     (bus_gic_sock)     {};
    \node[tsocki]                                      at ([xshift=-0.3cm]gic.north)       (gic_socki_0)      {};
    \node[tsocki]                                      at ([xshift=-0.1cm]gic.north)       (gic_socki_1)      {};
    \node[tsocki]                                      at ([xshift=+0.1cm]gic.north)       (gic_socki_2)      {};
    \node[tsocki]                                      at ([xshift=+0.3cm]gic.north)       (gic_socki_3)      {};
    \node[isocki]                                      at ([xshift=-0.3cm]gic.south)       (gic_socki_i)      {};
    \draw (gic_sock)      -- (bus_gic_sock);

    \node[nodetype, above=of bus.north, xshift=+1.4cm]                                     (uart)             {UART};
    \node[tsock]                                       at (uart.south)                     (uart_sock)        {};
    \node[isock]                                       at (uart_sock.center|-bus.north)    (bus_uart_sock)    {};
    \node[isocki]                                      at (uart.north)                     (uart_socki)       {};
    \draw (uart_sock)     -- (bus_uart_sock);

    \node[nodetype, below=of bus.south, xshift=-2.1cm]                                     (mem)              {MEM};
    \node[tsock]                                       at (mem.north)                      (mem_sock)         {};
    \node[isock]                                       at (mem_sock.center|-bus.south)     (bus_mem_sock)     {};
    \draw (mem_sock)      -- (bus_mem_sock);

    \node[nodetype, below=of bus.south, xshift=-0.7cm]                                     (sdhci)             {SDHCI};
    \node[tsock]                                       at ([xshift=+0.2cm]sdhci.north)     (sdhci_t_sock)      {};
    \node[isock]                                       at ([xshift=-0.2cm]sdhci.north)     (sdhci_i_sock)      {};
    \node[isock]                                       at (sdhci_t_sock.center|-bus.south) (bus_sdhci_i_sock)  {};
    \node[tsock]                                       at (sdhci_i_sock.center|-bus.south) (bus_sdhci_t_sock)  {};
    \node[isocki]                                      at (sdhci.south)                    (sdhci_socki)       {};
    \draw (sdhci_t_sock)  -- (bus_sdhci_i_sock);
    \draw (sdhci_i_sock)  -- (bus_sdhci_t_sock);

    \node[nodetype, below=of bus.south, xshift=+0.7cm]                                     (ethoc)            {ETHOC};
    \node[tsock]                                       at ([xshift=+0.2cm]ethoc.north)     (ethoc_t_sock)     {};
    \node[isock]                                       at ([xshift=-0.2cm]ethoc.north)     (ethoc_i_sock)     {};
    \node[isock]                                       at (ethoc_t_sock.center|-bus.south) (bus_ethoc_i_sock) {};
    \node[tsock]                                       at (ethoc_i_sock.center|-bus.south) (bus_ethoc_t_sock) {};
    \node[isocki]                                      at (ethoc.south)                    (ethoc_socki)      {};
    \draw (ethoc_t_sock)  -- (bus_ethoc_i_sock);
    \draw (ethoc_i_sock)  -- (bus_ethoc_t_sock);

    \node[nodetype, below=of bus.south, xshift=+2.1cm]                                     (rng)              {RNG};
    \node[tsock]                                       at (rng.north)                      (rng_sock)         {};
    \node[isock]                                       at (rng_sock.center|-bus.south)     (bus_rng_sock)     {};
    \draw (rng_sock)      -- (bus_rng_sock);

    \draw[draw=rwth] (cpu_socki)   -- +(0,+0.2cm) -| (gic_socki_0);
    \draw[draw=rwth] (uart_socki)  -- +(0,+0.2cm) -| (gic_socki_3);
    \draw[draw=rwth] (cpu_sockit)  -- +(0,-0.2cm) -| (gic_socki_i);
    \draw[draw=rwth] (sdhci_socki) |- ([xshift=+0.4cm, yshift=-0.4cm]sdhci_socki.center-|bus.east) |- ([yshift=+0.4cm]gic_socki_1.center) -- (gic_socki_1);
    \draw[draw=rwth] (ethoc_socki) |- ([xshift=+0.3cm, yshift=-0.3cm]ethoc_socki.center-|bus.east) |- ([yshift=+0.3cm]gic_socki_2.center) -- (gic_socki_2);
    
    \node[isock]                                        at ([yshift=-2.3cm]bus.west)                    (isock_legend)  {};
    \node[tsock]                                        at ([xshift=+0.5cm]isock_legend-|bus.center)    (tsock_legend)  {};
    \node[isocki, below=of isock_legend, yshift=+0.4cm]                                                 (isocki_legend) {};
    \node[tsocki]                                       at (isocki_legend-|tsock_legend)                (tsocki_legend) {};
    
    \node[draw=none, right=of isock_legend,  xshift=-0.5cm, anchor=west] (isock_legend_text)  {\footnotesize\acs{tlm} Initiator Socket}; 
    \node[draw=none, right=of tsock_legend,  xshift=-0.5cm, anchor=west] (tsock_legend_text)  {\footnotesize\acs{tlm} Target Socket}; 
    \node[draw=none, right=of isocki_legend, xshift=-0.5cm, anchor=west] (isocki_legend_text) {\footnotesize\acs{tlm} Interrupt Initiator Socket}; 
    \node[draw=none, right=of tsocki_legend, xshift=-0.5cm, anchor=west] (tsocki_legend_text) {\footnotesize\acs{tlm} Interrupt Target Socket}; 
\end{tikzpicture}
    \caption{\acs{avp64} architecture.}
    \label{fig:vp}
    \vspace{-1em}
\end{figure}

\cref{fig:vp} depicts the architecture of the \ac{vp}.
It consists of an ARMv8 CPU and peripherals that are connected via a bus.
Interrupts are implemented by a \ac{tlm}-based interrupt protocol.
An \ac{os} kernel like Linux can be booted from a virtual \ac{sd} card using the \ac{sdhci}.
PL011 ARM PrimeCell \acs{uart} models serve as user interface.
They can be configured to print their output to the host's terminal and read-in user input.
The \ac{vp} is a loosely-timed simulation, i.e., the SystemC threads are temporal decoupled, keep their local time, and regularly synchronize with the simulation time.
Memory accesses and interrupts are implemented by \ac{tlm} transactions.

All benchmarks have been executed on a computer equipped with a \textit{\acs{amd} Ryzen 9 3900X 12-Core} CPU, \SI{64}{\giga\byte} \acs{ram}, and a \textit{Samsung 860 EVO \acs{sata} III} \acs{ssd}, running Cent\acs{os}~7.9 with Linux~3.10.0.
The maximum allowed quantum duration for the simulation was \SI{100}{\micro\second}.

\subsection{Case Study}
\label{sec:exp-eval:case-study}
\begin{figure}[!htb]
    \vspace{-1em}
    \centering
    \subfloat[Real-time factor.]{    
        \includegraphics[width=\linewidth]{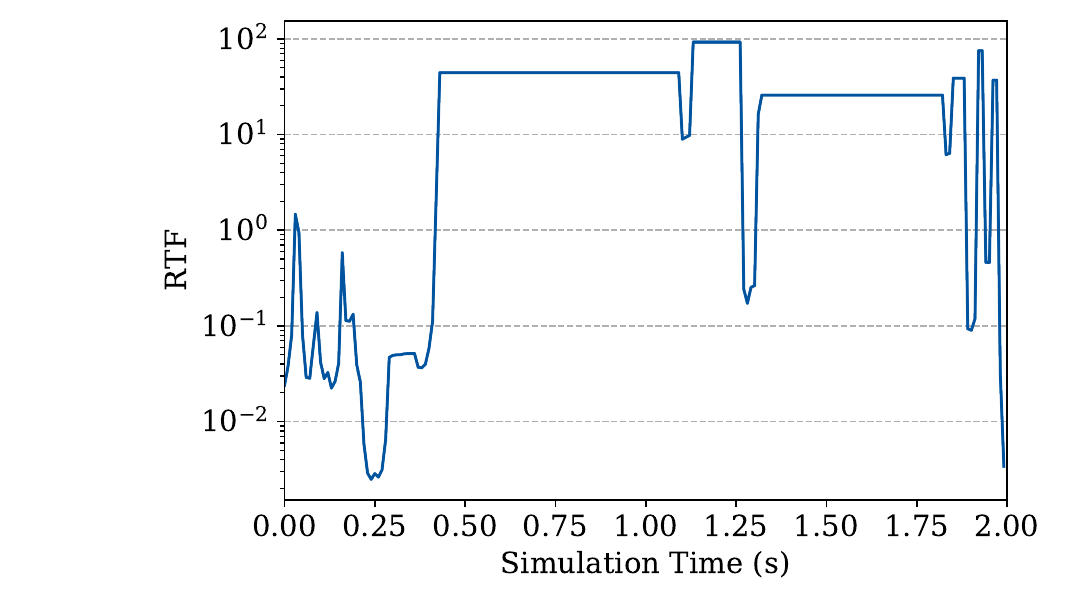}
        \label{fig:linux-boot-rtf}
    }
    \hfill
    \subfloat[Quantum durations of the processor thread.]{
        \includegraphics[width=\linewidth]{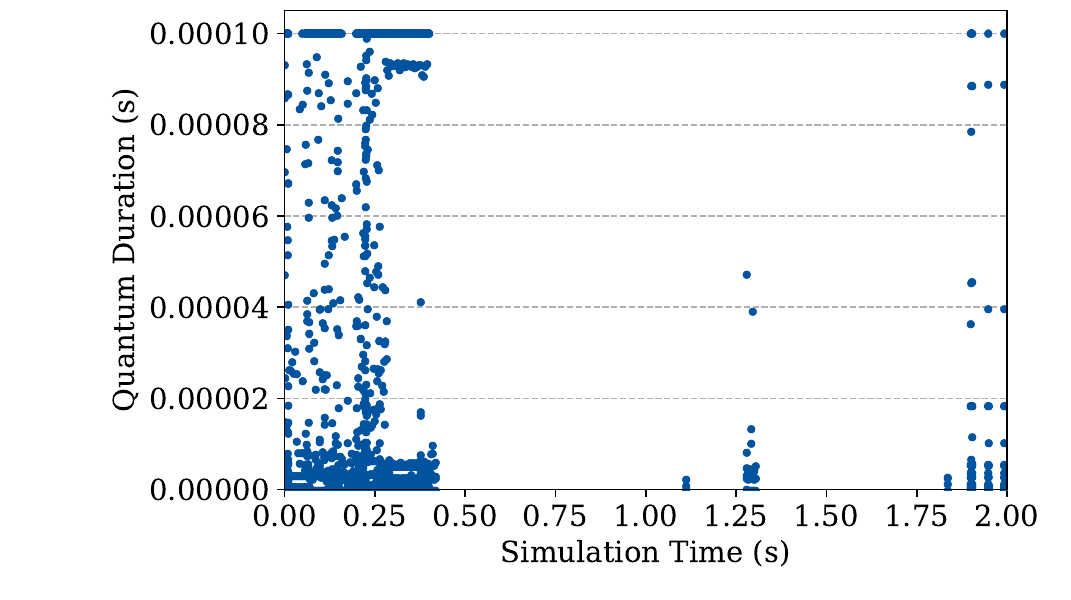}
        \label{fig:linux-boot-quantum}
    }
    \hfill
    \subfloat[Excerpt of notified SystemC events.]{    
        \includegraphics[width=\linewidth]{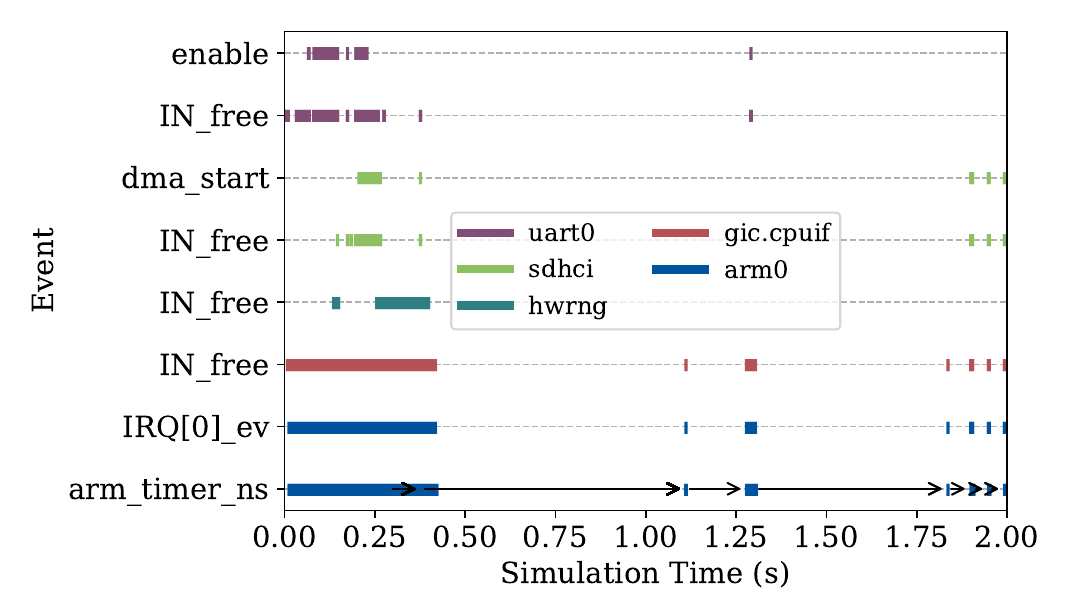}
        \label{fig:linux-boot-event}
    }
    \caption{Tracing results of the first \SI{2}{\second} of Linux boot.}
    \label{fig:linux-boot}
\end{figure}

\acs{avp64} was started with \nistt being preloaded.
\cref{fig:linux-boot} depicts the captured results of the first \SI{2}{}~seconds of the Linux boot process.
The \ac{rtf} during the simulation is shown in \cref{fig:linux-boot-rtf}.
It can be observed that the \ac{rtf} fluctuates during the simulation.
This fluctuation depends on the workload that is executed on the \ac{vp}.
Different parts of the workload cause different CPU utilization.
When the utilization is low, idle cycles of the CPU are not simulated, which increases the performance.
Besides that aspect, interactions with peripherals can also reduce the \ac{rtf} due to early quantum terminations.

\cref{fig:linux-boot-quantum} shows the used quantum durations of the CPU thread.
During the periods where the \ac{rtf} is low, the quanta are comparatively small.
If no data points are printed, that point in time has not been simulated for the CPU model.
The reason for that is that the state of the CPU stayed unchanged during that time.
That is, e.g., the case when the CPU is in idle mode.
The Linux idle mode is implemented by executing the \ac{wfi} instruction.
\acs{avp64} emulates this instruction by using the SystemC \texttt{wait} function with the interrupt event as a parameter.
Thereby, the execution of the processor thread is suspended until the next interrupt raises.

\cref{fig:linux-boot-event} visualizes the notified SystemC events.
The \texttt{IN\_FREE} event is used by the \ac{vcml} register model to serialize parallel accesses to \ac{tlm} target sockets of a peripheral.
It is notified after each handled transaction of the corresponding peripheral.
That means, every time the \texttt{IN\_FREE} event is notified, a register of the peripheral has been written.
Thus, the figure reveals when interactions with peripherals take place.

Besides the \texttt{IN\_FREE} event, the events \texttt{IRQ[0]\_ev} and \texttt{arm\_timer\_ns} of the CPU model are of importance.
The first event, \texttt{IRQ[0]\_ev}, is notified every time an interrupt is signaled to the CPU by the \ac{gic}.
That is the reason why the notification pattern of the \texttt{IRQ[0]\_ev} of the CPU and the \texttt{IN\_FREE} event of the \ac{gic} look similar.
Every time an interrupt is signaled to the CPU, the CPU interacts with the CPU interface of the \ac{gic} to handle and acknowledge the interrupt.
The second event of the CPU, \texttt{arm\_timer\_ns}, is used to implement a timer.
When the timer is set to expire after a certain time, the event is programmed to be notified at this time.
This delayed notification is visualized in \cref{fig:linux-boot-event} by an arrow starting at the time of programming and pointing to the time of expiration.
There are three periods during the boot process where the timer is programmed to expire after a comparatively long period of time (\SI{0.4}{\second}-\SI{1.1}{\second}; \SI{1.1}{\second}-\SI{1.3}{\second}; \SI{1.3}{\second}-\SI{1.8}{\second}).
During these periods, no quantum data were recorded.
That strengthens the assumption that the \ac{os} was in idle mode during those periods and used the timer to wake up.

\Cref{tab:linux-boot-tot-time-per-proc} shows the needed computation time to execute the \texttt{SC\_THREAD}s of the \ac{vp}.
The simulation of \SI{2}{\second} of the Linux boot took \SI{29.15}{\second}.
\SI{96}{\percent} of the simulation, the processor thread of the CPU model was active.
Since the boot process is used as the benchmark, the CPU was the most compute-intensive model of the platform.

\begin{table}[!htb]
    \centering
    \caption{SystemC thread execution time during \SI{2}{\second} of Linux boot.}
    \label{tab:linux-boot-tot-time-per-proc}
    \begin{tabular}{l l S[table-format=2.5] S[table-format=2.6]}
        \toprule
        Module                 & SC\_THREAD        & {Compute Time (\si{\second})} & {Amount (\si{\percent})} \\ \midrule
        System                 & timeout           &  0.000001                     &  0.000003                \\ \hdashline
        \multirow{2}{*}{ETHOC} & tx\_process       &  0.002494                     &  0.008556                \\ \cdashline{2-3}
                               & rx\_process       &  0.002670                     &  0.009160                \\ \hdashline
        SDHCI                  & dma\_tread        &  0.239657                     &  0.822207                \\ \hdashline
        CPU                    & processor\_thread & 27.978902                     & 95.989095                \\ \midrule 
        Total                  &                   & 28.223724                     & 96.829022                \\ \bottomrule
    \end{tabular}
\end{table}

\subsection{Performance}
\label{sec:exp-eval:performance}
The idea of \nistt is to trace the execution of a \ac{vp} and to place as few requirements as possible on the simulation.
Therefore, \ldpreload is used to limit the requirements to the usage of a shared SystemC library.
This section examines whether the performance of \nistt is reduced due to its non-intrusive implementation.
If changing simulation-dependent libraries had not conflicted with the requirements, instrumentation of functions of interest inside the SystemC library would have been an alternative implementation with the same tracing results as \nistt.
This alternative, intrusive implementation is compared to \nistt to evaluate the differences in performance. 
Furthermore, the general overhead of \nistt is classified.

The simulation of the \ac{vp} is executed in four different configurations to evaluate the tracing overhead of \nistt.
As shown in \cref{tab:tracing-cfg}, the configurations differ in the tracing implementation and the way the simulation is linked against SystemC.
\nistt requires a dynamic linkage against SystemC to work.
The intrusive reference implementation is used to profile the overhead of intercepting function calls using preloading.
As a reference, the execution time of the unmodified \ac{vp} without tracing is measured.

\begin{table}[!htb]
    \centering
    \caption{Tracing implementations.}
    \label{tab:tracing-cfg}
    \begin{tabular}{c c c}
        \toprule
        Name               & Tracing        & SystemC Linkage \\ \midrule
        Reference          & -              & dynamic         \\ \hdashline
        I,static           & intrusive      & static          \\ \hdashline
        I,shared           & intrusive      & dynamic         \\ \hdashline
        NI,shared (\nistt) & non-intrusive  & dynamic         \\ \bottomrule
    \end{tabular}
\end{table}

\begin{figure}[!htb]
    \centering
    \includegraphics[width=\linewidth]{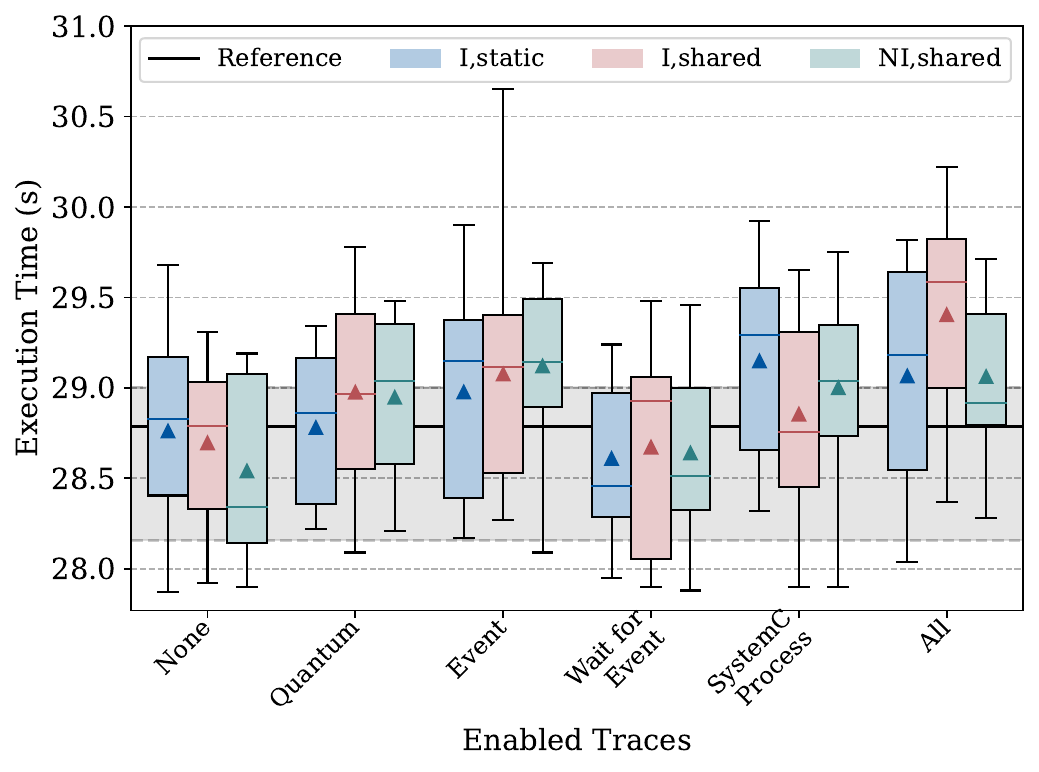}
    \caption{Overhead of tracing during Linux boot simulation for the intrusive (I) and non-intrusive (NI) implementations. Results of $20$ runs.}
    \label{fig:overhead}
\end{figure}

The first \SI{2}{\second} of the Linux boot process are again used as the benchmark to quantify the overhead of \nistt and the two intrusive implementations compared to the reference \ac{vp} without tracing.
\cref{fig:overhead} shows the results for the different configurations and different activated traces.

When the tracing is implemented but deactivated (cf. \textit{None}), i.e., additional functions are present in either SystemC or the preloaded \nistt library, but data are not stored in a database, the execution times are similar to the one of the reference implementation without tracing.
In general, no clear relation between used linking and needed execution time could be detected.
Even \nistt does not harm the simulation performance compared to the corresponding intrusive implementation.
What is noticeable, however, is that the kind of tracing that is enabled has an influence on the execution time.
That is because the traces are executed with different frequencies and therefore cause various overheads.
The \textit{Wait for Event} trace, e.g., is only triggered \SI{68}{} times during the simulation and therefore caused only a little overhead.
The \textit{Quantum} trace captured \SI{94174}{} data points which results in a visibly higher execution time.
The highest overhead is created, and most data points are stored, by the \textit{Event} trace (\SI{195064}{} data points) and the \textit{SystemC Process} trace (\SI{208174} data points).
Those results show that the produced overhead is mainly caused by the kind of the trace rather than its implementation (intrusive/non-intrusive) or used linkage.
However, a non-intrusive implementation like \nistt has the advantage that the simulation does not need to be adapted to the tool.
The standardized SystemC \ac{api} assures compatibility.
Furthermore, no access to the source code of the simulation is needed for the non-intrusive implementation.
The simulation itself stays unchanged and does not need to be recompiled to be traced.

\section{Conclusion}
\label{sec:conclusion}
This paper presents a novel approach for tracing a SystemC-\acs{tlm}-2.0-based simulation in a non-intrusive way.
Due to the standardization of the SystemC \ac{api}, \nistt can trace every simulation that is based on SystemC without making special requests to the implementation.
Source code or debug symbols of the simulation are not needed, which drastically increases the usability compared to existing approaches.
\nistt stores the captured tracing data in a database to allow evaluation and analysis in a post-processing step.
We showed that the non-intrusive implementation of \nistt does not reduce the performance compared to an intrusive one.
The non-intrusive implementation has the advantage of not requiring compile-time modifications.
Thereby, also \acp{vp} can be analyzed where the source code can not be accessed.

However, there is also a limitation of \nistt due to its implementation using preloading.
\nistt is only capable of redirecting calls to SystemC functions that are located in the library object file.
Thereby, calls to inlined functions or functions of template classes cannot be intercepted.
Despite this limitation, \nistt is a powerful tool that offers deep insights into SystemC-\acs{tlm}-2.0-based simulations like \acp{vp} without the need of having access to their source code.
It can capture relevant \ac{esl}-simulation data and can easily be extended to trace additional functions of interest.

\FloatBarrier

\bibliographystyle{IEEEtran}
\bibliography{IEEEabrv,bib/Library}
\vspace{12pt}

\end{document}